\documentclass[doublecol]{epl2}

\title{Striped antiferromagnetism and electronic structures of SrFeAsF and their implications}
\shorttitle{Striped antiferromagnetism and electronic structures of SrFeAsF and their implications}

\author{Li-Fang Zhu\inst{1,2} \and Bang-Gui Liu\inst{1,2}}
\shortauthor{Li-Fang Zhu and Bang-Gui Liu}

\institute{ \inst{1} Institute of Physics, Chinese Academy of
Sciences, Beijing 100190, China\\
\inst{2}Beijing National Laboratory for Condensed Matter Physics,
Beijing 100190, China}


\abstract{We investigate structural, magnetic, and electronic
properties of SrFeAsF as a new parent for superconductors using a
state-of-the-art density-functional theory method. Calculated
results show that the striped antiferromagnetic order is the
magnetic ground state in the Fe layer and the interlayer magnetic
interaction is tiny. Calculated As and Sr positions are in good
agreement with experiment. There are only two
quasi-two-dimensional bands near the Fermi level. The valence
charge is mainly in the Fe and F layers, and the magnetic moment
is confined to the Fe atoms. All the spin couplings within the Fe
layer are antiferromagnetic due to the superexchange through the
nearest As atoms. These results, with the record-equaling
phase-transition temperature in the latest Sm-doped SrFeAsF, show
that the SrFeAsF, sharing the same structure with LaFeAsO, is
promising for achieving better superconductors.}

\begin{document}

\maketitle

\section{Introduction}

Fe-based superconductors attract more and more attention since
superconductivity was found in doped LaFePO\cite{LaOFeP}. The
advent of superconducting F-doped LaFeAsO stimulates a world-wide
campaign for more and better Fe-based
superconductors~\cite{SC-LaOFeAs}. Replacing La by other
lanthanides or doping with F has yielded more superconductors, and
higher phase-transition temperatures ($T_c$) have been achieved in
some of them~\cite{SC-LaOFeAs-n,SC-SmOFeAs-n,SC-SmOFeAs-zhao-55K}.
Furthermore, much more superconducting materials were found by
using other dopants and other parent
compounds~\cite{Peter,BaFe2As2-ES,SrFe2As2,EuFe2As2}. Even
$\alpha$ FeSe can be made superconducting by applying high
pressure~\cite{FeSe-27K,FeSe-pressure}. Now there are three series
of FeAs-based superconductors: $R$FeAsO ($R$: lanthanide
elements), $A$Fe2As2 ($A$: alkaline-earth elements), and LiFeAs.
So far, the highest $T_c$ is 55-56 K in the case of doped
SmFeAsO~\cite{SC-SmOFeAs-zhao-55K}. Their structural, magnetic,
electronic properties are intensively investigated and the
microscopic mechanism for the superconductivity in these materials
has been
explored~\cite{SC-LaOFeAs-n-MO,SC-LaOFeAs-n-MO1,add,LaOFeAs-ES,LaOFeAs-AFM,LaOFeAs-150K,LaOFeAs-ES-SDW,LaOFeAs-OP,LaOFeAs-EC,moment,mazinprb,caoprb}.
Very recently, superconductivity was found in Co and La doped
SrFeAsF materials~\cite{hosono,wen2,SrFeAsF-56K,SrFeAsF-7}.
SrFeAsF has the same crystal structure as $R$FeAsO and similar
magnetic instability, but does not include any
lanthanide~\cite{epl,wen1,SrFeAsF-muon}. It has been found that
Sm-doped SrFeAsF can become superconducting at 56 K, and higher
transition temperatures should be reached with appropriate
dopants~\cite{wen2,SrFeAsF-56K,SrFeAsF-7}. Because some magnetic
fluctuations are believed to mediate the superconductivity in
FaAs-based materials, it is highly desirable to investigate the
magnetic orders, electronic structures, and magnetic properties of
the parent compound SrFeAsF.

Here, we use a state-of-the-art density-functional theory (DFT)
method to investigate the structural, electronic, and magnetic
properties of the SrFeAsF. Our total energy results show that the
striped antiferromagnetic (AF) order, the same as that of LaFeAsO,
is the magnetic ground state in the Fe layer, and the interlayer
magnetic interaction is tiny. Our calculated position parameters
of As and Sr are in good agreement with experiment. The electronic
band result shows that there are only two quasi-two-dimensional
(quasi-2D) bands near the Fermi level. Our charge and
magnetization density analysis shows that the valence charge is
mainly distributed in the Fe and F layers, and the magnetic moment
is confined to the Fe layer. The spin couplings within the Fe
layer are AF due to superexchange through the nearest As atoms.
The real moment should be much smaller than the DFT value because
of quantum spin fluctuations. More detailed results will be
presented in the following.

\section{Computational detail}

\begin{figure}
\includegraphics[width=8.8cm]{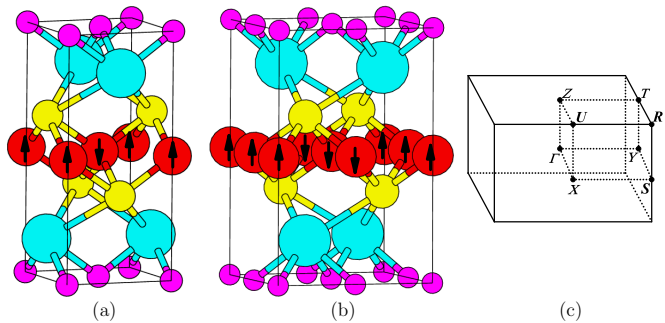}
\caption{(color online). The checkerboard
structure [AF1, (a)], the stripe one [AF2, (b)], and the first
Brillouin zone of the AF2 structure (c). The largest ball (cyan or
white) denotes Sr, the medium (yellow or white) As, the smallest
(magenta or gray) F, and the ball with arrow Fe (red or gray). The
arrow denotes the spin orientation.}\label{fig.1}
\end{figure}

SrFeAsF usually assumes the tetragonal AsCrSiZr (tP8) structure
with space group P4/nmm (No. 129) at high temperature. It transits
to an orthorhombic distorted phase at 185 K, and furthermore,
there is a magnetic instability at 175 K~\cite{epl,wen1}. A
similar behavior has been observed in
LaFeAsO~\cite{SC-LaOFeAs,SC-LaOFeAs-n-MO,SC-LaOFeAs-n-MO1}. We
shall investigate by self-consistent calculations the possible
magnetic structures of the two SrFeAsF phases shown in
fig.~\ref{fig.1}. Figure~\ref{fig.1}a shows the structure with the
familiar checkerboard AF (AF1) order and fig.~\ref{fig.1}b that
with a striped AF (AF2) order. The interlayer magnetic interaction
can be ferromagnetic (FM) or AF.

All the calculations are done using a full-potential
linearized-augmented-plane-wave (FLAPW) method within the
density-functional theory~\cite{dft1,dft2}, as implemented in the
package WIEN2K~\cite{wien2k,wien2ka}. The generalized gradient
approximation (GGA) to the exchange-correlation potential is used
for the presented results~\cite{pbe96}, and
local-density-approximation (LDA) calculations are done for
comparison~\cite{lda}. Full relativistic effects are calculated
for core states, and the scalar relativistic approximation is used
for valence states. The spin-orbit coupling ~\cite{relsa} is
neglected because it has little effect on the system. We use 400 k
points in the first Brillouin zone for the two AF structures. We
make the harmonic expansion up to $l_{\rm max}$=10 inside the
atomic spheres. The radii of the muffin-tin spheres of Sr, Fe, As,
and F are 2.3, 2.1, 2.2 and 2.0 atomic unit (a.u.), respectively.
$R_{\rm mt}$$\times$$K_{\rm max}$ is set to 8.0. The
self-consistent calculations are considered to be converged only
when the integrated charge difference per formula unit between
input and output charge density is less than 0.0001.

\begin{figure}
\includegraphics[width=8.8cm]{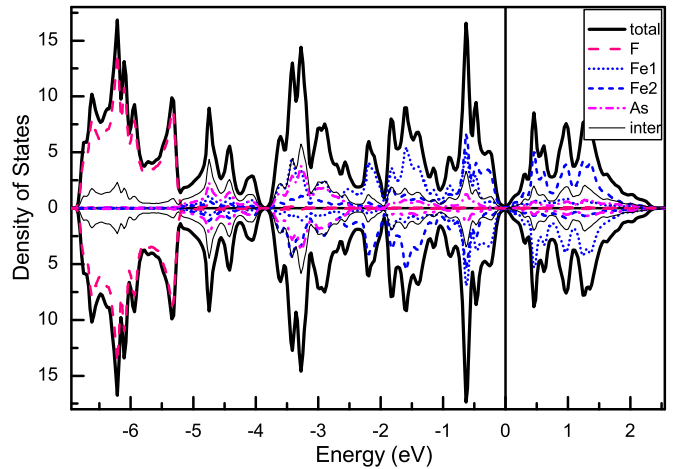}
\caption{(color online). Total density of
states (black thick solid, in states/eV per formula unit) and
those projected in atomic spheres of F (pink or gray dash), Fe1
(blue or gray dot), Fe2 (blue or gray short dash), and As (magenta
or gray dash dot) and the interstitial region (black thin sold) of
the AF2 structure.}\label{fig.2}
\end{figure}

\section{Results and discussion}

We use the experimental lattice constants ($a$,$c$) =
(3.9930\AA{},8.9546\AA{}) for the tetragonal structure, and
($a$,$b$,$c$) = (5.6155\AA{},5.6602\AA{},8.9173\AA{}) for the
orthorhombic structure~\cite{epl,wen1}. As for the internal
position parameters of As and Sr atoms, we use ($z_{\rm
As}$,$z_{\rm Sr}$) = (0.6527,0.1598) and (0.6494,0.1635) as input
and optimize them in terms of forces standard (2 mRy/a.u.) for
both of the structures~\cite{epl,wen1}. We obtain ($z_{\rm
As}$,$z_{\rm Sr}$) = (0.6444,0.1634) and (0.6475,0.1637) for the
AF1 and AF2 structures, respectively. For the AF2 structure, our
GGA result, (0.6475,0.1637), is in good agreement with
experiment~\cite{epl,wen1}, in contrast to that from non-magnetic
calculations~\cite{nekrasov,shein} and (0.6357,0.1624) of an LDA
calculation of ours. The magnetic moment in the Fe sphere is 1.65
and 1.97$\mu_B$ for the AF1 and AF2 phases. The total Fe moment in
the AF2 phase is approximately 2$\mu_B$, twice as large as an LDA
result of ours. The parameters $z_{\rm As}$ and $z_{\rm Sr}$ and
the magnetization density distributions do not change when we
switch the interlayer magnetic interaction from FM to AF. The AF2
phase is lower by 80 meV in total energy per formula unit than the
AF1 phase. This means that the striped AF2 phase is the
ground-state phase, in agreement with experiment~\cite{epl,wen1}.
The Fe spins align parallel along the $a$ (shorter) axis and
antiparallel along the $b$ axis, which is the same as
LaFeAsO~\cite{Peter,LaOFeAs-150K,mazin}. SrFeAsF has the striped
AF order as its ground state, the same as
LaFeAsO~\cite{LaOFeAs-AFM,LaOFeAs-150K,mazinprb,caoprb}. The
energy difference of 80 meV of the AF1 phase is also comparable
with the 75 meV for the checkerboard AF phase of
LaFeAsO~\cite{LaOFeAs-AFM}. The moment of the magnetic ground
state phase is approximately the same as that of
LaFeAsO~\cite{LaOFeAs-AFM,mazinprb,caoprb}.

\begin{figure}
\includegraphics[width=8.8cm]{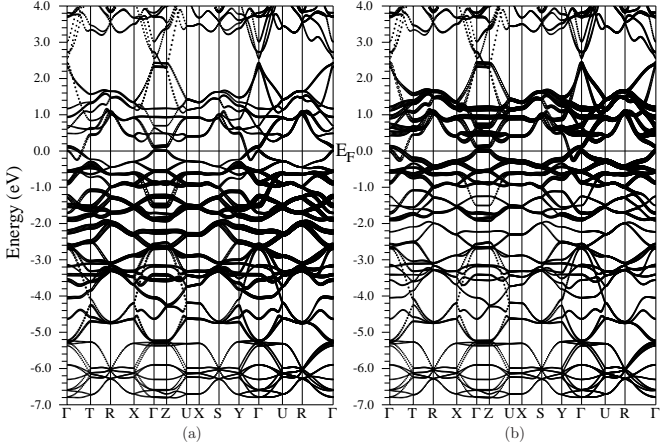}
\caption{Spin-dependent band structure of the
AF2 structure along representative high-symmetry lines in the
first Brillouin zone. The left part is that with Fe1 spin-up
character shown and the right part with Fe1 spin-down character
shown. The band consists of dots, where the bigger the dot, the
stronger Fe1 character is at that point.}\label{fig.3}
\end{figure}

The total spin-dependent density of states (DOS) and those
projected in the muffin-tin spheres of F, Fe1, Fe2, and As atom
and in the interstitial region of the AF2 structure are presented
in fig.~\ref{fig.2}. The filled states between -6.9 and 0 eV are
the valence states from Fe-3d$^6$4s$^2$, As-4p$^3$, Sr-5s$^2$, and
F-2p$^5$ orbitals. The semi-core states such as As-4s$^2$ are
lower than -10.7 eV. There is no energy gap in the energy window
presented, but two pseudo-gaps are visible, at the Fermi level and
-3.85 eV. The DOS between -6.9 and -5.2 eV, having symmetry
between the two spin channels, mainly comes from F-2p states, and
the Sr states almost disappear from the energy window presented.
The As DOS is also almost symmetrical. Concerning spin, it is
distributed between -5.2 and 2.5 eV, but its largest part is
between -3.8 and -2.2 eV. It is clear that the Fe states are
spin-split. Actually, we have two different Fe atoms: Fe1 and Fe2.
Their spins are antiparallel. The spin-down part of the Fe1 DOS is
mainly distributed between -1 and 2.5 eV and the spin-up part
between -3.8 and -1 eV. The Fe2 DOS is equivalent to the Fe1 DOS
with the two spin channels interchanged. One feature is a
substantial DOS contribution from the interstitial region, which
reflects substantial covalence between Fe and As. At the Fermi
level, the DOS calculated without spin polarization is
substantially larger than that of our spin-polarized
calculation~\cite{nekrasov,shein}. Our DOS results are similar to
spin-polarized GGA results of LaFeAsO~\cite{LaOFeAs-AFM}.

\begin{figure}
\includegraphics[width=8.8cm]{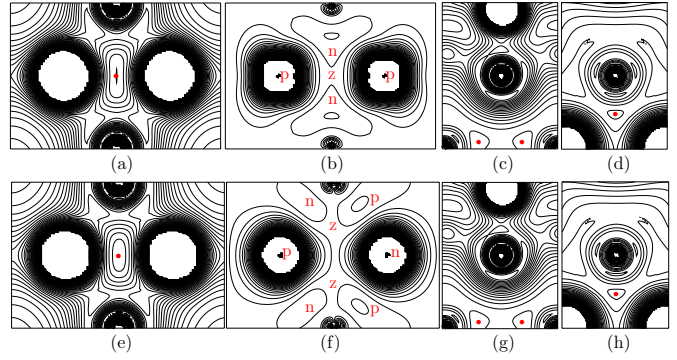}
\caption{(color online). The charge density
distributions in the two Fe-As-Fe planes with the Fe atoms along
the $x$ axis (a) and the $y$ axis (e), the two different Sr-As-Sr
planes (c) and (g), and the two F-Sr-F planes with the F atoms
along the $x$ axis (d) and the $y$ axis (h); and the magnetization
density distributions in the same two Fe-As-Fe planes (b) and (f).
The charge density increment is 0.003$e$/a.u.$^3$ and the
magnetization one $\pm$0.003$\mu_B$/a.u.$^3$. The red or gray dot
in the charge plots labels the smallest density. In the
magnetization plots, the red or gray letters `z', `p', and `n'
mean that the magnetization density therein is zero, positive, and
negative, respectively.} \label{fig.4}
\end{figure}

In fig.~\ref{fig.3} we show the spin-dependent band structure with
Fe1 character. The bands have very little dispersion along the $z$
axis. It is surprising that the band structure near the Fermi
level is very simple, consisting of only two bands at the Fermi
level. The band that has the maximum along $\Gamma$-Z originates
from the Fe1 d$_{xy}$ state, and the other the Fe1 d$_{x^2-y^2}$
one. These two bands near the Fermi level have a much larger
weight in the spin-down than in the spin-up channel. This implies
that the electronic structure near the Fermi level is quasi-2D and
spin-split. This special property is reasonable because the Sr
atom is strongly ionic and the Fe moments align
antiferromagneticaly along the $y$ axis. The band structure near
the Fermi level is similar to that of LaFeAsO~\cite{LaOFeAs-AFM}.
It is interesting that the band feature near the Fermi level along
$\Gamma$-$Y$ is consistent with that obtained with a
phenomenological two-band model~\cite{leedh}. Actually, the
detailed band structure near the Fermi level depends sensitively
on the structural parameters ($z_{\rm As}$,$z_{\rm Sr}$) and the
magnetic moment~\cite{LaOFeAs-AFM,mazinprb,caoprb}. When $z_{\rm
As}$ becomes smaller, the moment is correspondingly smaller and
the band structure near the Fermi level looks more like that of
non-magnetic
calculations~\cite{nekrasov,shein,LaOFeAs-ES-SDW,mazinprb}.

\begin{figure}
\includegraphics[width=8.8cm]{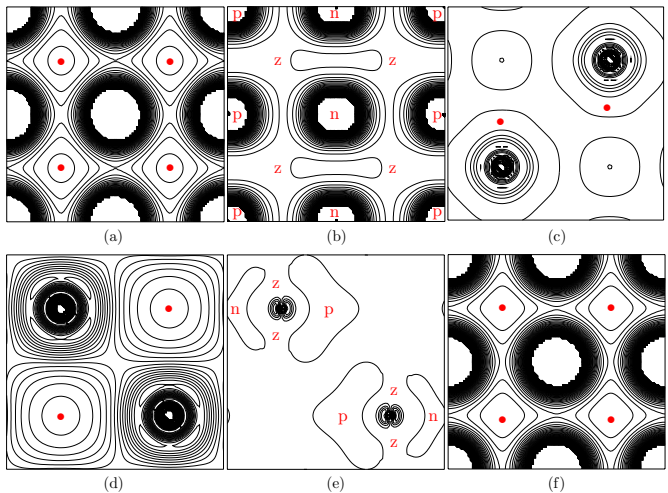}
\caption{(color online). The charge density
distributions in the Fe plane (a), the Sr plane (c), the As plane
(d), and the F plane (f); and the magnetization density
distributions in the Fe plane (b) and the As plane (e). The red or
gray dot labels the smallest density. The red or gray letters `z',
`p', and `n' and the density increments are the same as in Fig.
4.}\label{fig.5}
\end{figure}

In fig.~\ref{fig.4} we present the charge and magnetization
density distributions in the planes defined by the two symmetrical
bonds with the same vertex atom. The panels (a) and (e) show the
charge density distributions in the two planes defined by the two
different Fe-As-Fe chains, respectively, and the (b) and (f) the
corresponding magnetization density distributions. One of the
planes includes the $x$ axis and the other the $y$ axis. The
panels (c) and (g) show the charge density distributions in the
two planes defined by the two different Sr-As-Sr chains. The
panels (d) and (h) show the charge density distributions in the
two planes defined by the two different F-Sr-F chains,
respectively. The red or gray dot in the charge plots labels the
smallest charge density in $e$/a.u.$^3$ ($e$ is the electron
charge): 0.007 in (a), 0.003 in (c), 0.003 in (d), 0.006 in (e),
0.003 in (g), and 0.003 in (h). The lowest contours represent the
charge density values in $e$/a.u.$^3$: 0.004, 0.007, 0.010, and so
on. In the magnetization plots in (b) and (f), the red or gray
letters `z', `p', and `n' mean that the magnetization density in
the region is zero, positive, and negative, respectively. The
magnetization density values (in $\mu_B$/a.u.$^3$) of the contours
near the zero are $\pm 0.003$, $\pm 0.006$, and so on. The
magnetization density vanishes in these planes at the same places
as in the panels (c), (d), (g), and (h), and thus this is not
shown again.

In order to investigate the charge and magnetization density
distributions as a whole, we present them by atomic layer in
fig.~\ref{fig.5}. The panels (a) and (d) show the charge density
distributions in the Fe and As planes, and the panels (b) and (e)
the magnetization density distributions in the same planes. The
panels (c) and (f) show the charge density distributions in the Sr
and F planes, and the magnetization density in these two planes is
near zero and thus not presented. The red or gray dots denote the
smallest charge density in $e$/a.u.$^3$: 0.024 in (a), 0.0025 in
(c), 0.006 in (d), and 0.0025 in (f). The letters `z', `p', and
`n' imply the same as in fig.~\ref{fig.4}. It is clear that the
spin is mainly confined in the Fe layer and the magnetization
density is nearly zero in the F layer and the adjacent two Sr
layers. Therefore, the interlayer spin interaction must be tiny,
which in return supports our computational model.

In addition, we have done the same DFT calculations assuming an AF
order along the $z$ axis. The calculated total energies and
position parameters of As and Sr are the same as those presented
above. This is caused by the strong ionicity of the Sr layers and
the zero magnetization density in the Sr and F layers. Actually,
one cannot determine the magnetic order along the $z$ axis by
using DFT calculations because the interlayer magnetic interaction
is too weak. As for the spin order in the Fe layer, the spins
align antiferromagnetically along the $y$ axis but
ferromagnetically along the $x$ axis. The nearest spin coupling
constants along the $x$ and $y$ axes, $J_x$ and $J_y$, are AF.
There is an AF coupling $J^\prime$ ($>J_x/2$) between the next
nearest Fe spins. Other inter-spin couplings are substantially
smaller~\cite{zlf}. These three coupling constants are determined
by the super-exchange through the bridging As atoms and other
factors~\cite{mazin}. The small difference $\delta=J_y-J_x$ is
caused by the structural distortion.
 As for the moment per Fe atom, our GGA and LDA
results are nearly 2$\mu_B$ and 1$\mu_B$, much larger than the
experimental value $\sim 0.3\mu_B$~\cite{SrFeAsF-muon}. This
situation is the same as that of LaFeAsO whose experimental moment
per Fe atom is $0.25\sim 0.36$
$\mu_B$~\cite{SC-LaOFeAs-n-MO,SC-LaOFeAs-n-MO1}, much smaller than
$1.5\sim 2.3$ $\mu_B$ from GGA
calculations~\cite{mazinprb,caoprb}. For LaFeAsO, the large
discrepancies can be reduced by using LDA rather than GGA in
optimizing the crystal structure and calculating the moment, but
this scheme leads to a large discrepancy of the calculated As
position parameter from the experimental
result~\cite{LaOFeAs-150K,mazinprb,caoprb}. On the other hand,
spin-orbit coupling, monoclinic distortion, and p-d hybridization
are used to obtain smaller value for the moment~\cite{moment}. In
fact, this situation, however, has not been settled even in the
case of LaFeAsO. We believe that quantum many-body effects may
play some important roles in determining the actual moment. For
SrFeAsF, the actual magnetic moment is also much smaller than the
GGA value and we also attribute this discrepancy to quantum
many-body effects of the spin fluctuations. More investigations
are in need to solve this moment problem without leading to other
discrepancies.

\section{Conclusion}

In summary, we have investigated the structural, magnetic, and
electronic properties of SrFeAsF using the state-of-the-art DFT
method. Our total energy results show that the striped AF order is
the magnetic ground state in the Fe layer and the interlayer
magnetic interaction is tiny. The calculated As and Sr positions
are in good agreement with experiment. There are only two quasi-2D
bands near the Fermi level. The valence charge is mainly
distributed in the Fe and F layers, and the magnetic moment is
confined to the Fe atoms. All the intra-layer spin couplings are
AF due to the superexchange through As atoms. SrFeAsF shares the
main features of the structural, electronic, and magnetic
properties with LaFeAsO. The actual moment is smaller than our
value, which is the same as the situation for LaFeAsO and should
be solved by considering quantum many-body effects. Because the
Sm-doped SrFeAsF has equaled the $T_c$ record (56 K) of the
FeAs-based superconductors, it is believed that higher $T_c$ will
be realized in the SrFeAsF series. These results are useful for
understanding the structural, electronic, and magnetic properties
of SrFeAsF and should have implications achieving better
superconductors by appropriate doping.

\acknowledgments BGL is grateful to H. H. Wen for informing him of
the SrFeAsF series\cite{hosono,wen2,wen1,epl}. This work is
supported by Nature Science Foundation of China (Grant Nos.
10774180 and 10874232), by Chinese Department of Science and
Technology (Grant No. 2005CB623602), and by the Chinese Academy of
Sciences (Grant No. KJCX2.YW.W09-5).

\end{document}